\documentclass{elsart}

\usepackage{graphicx}
\usepackage{epsfig}

\usepackage{amssymb}

\begin{document}

\begin{frontmatter}

\title{Two-photon decays of $\rho$ and $\omega$ mesons and strong decays $\rho \to \eta \pi$
and $\omega \to \pi \pi$ in dense matter}

\author{A.E.Radzhabov}\ead{aradzh@thsun1.jinr.ru}\,
and \author{M.K.Volkov}\ead{volkov@thsun1.jinr.ru}

\address{Bogoliubov Laboratory of Theoretical Physics, \\
Joint Institute for Nuclear Research, 141980 Dubna, Russia}

\begin{abstract}
Two-photon decays of $\rho$ and $\omega$ mesons and strong decays $\rho \to \eta \pi$
and $\omega \to \pi \pi$ are described in the quark Nambu--Jona-Lasinio(NJL) model in
dense medium. Decays of $\rho$-meson are especially interesting because they contain
intermediate $a_0$-meson which became a sharp resonance in the vicinity of phase
transition of hadron matter into quark-gluon plasma. This leads to increasing
$\rho$-meson decay widths and, as a result, two-photon decay of $\rho$-meson in medium
became comparable with two-photon decays of scalar and pseudoscalar mesons in vacuum.
Similar effect take place in the strong decay $\rho \to \eta \pi$.
\end{abstract}

\end{frontmatter}

In the last years great attention is devoted to the study of the behavior of matter at
extreme conditions. These conditions can be reached in the heavy-ion collision
experiments which are performed now
\cite{Lourenco:2001wi,Roland:2001me,Gazdzicki:2004ef} and planned in future (SIS-300,
LHC). Indeed, at high temperatures and/or densities which can appear in these
experiments hadron matter can transit into quark-gluon plasma. The quark condensate
and spontaneous breaking of chiral symmetry disappear and the chiral symmetry is
restored. Let us emphasize that in this region many properties of elementary particles
are noticeably changed. Especially it concerns the scalar mesons, their masses
strongly decrease and approach the masses of the pseudoscalar mesons. The lifetime
changes and strong decay channels can be closed. As a result, broad scalar resonances
turn into sharp resonances. This can lead to the amplification of some processes
mediated by the scalar resonances, such as $\pi \pi \to \gamma \gamma$
\cite{Volkov:1997dx} and $\pi \pi \to \pi \pi$ \cite{Jido:2000bw}.

Especially interesting is that in this domain new decay modes which are forbidden in
vacuum can be opened up. It is two photon decays of vector mesons, strong decays $\rho
\to \eta \pi$ and $\omega \to \pi \pi$, dilepton decays of scalar mesons and so on.
This short paper is devoted to the description of the above mentioned vector meson
decays.

For description of these processes we use the $U(3)\times U(3)$ NJL model with 'tHooft
interaction \cite{Volkov:1986zb,Klimt:1989pm,Klevansky:1992qe,Volkov:1993jw,Ebert:1994mf}.
Lagrangian of the model consists of two parts: $U(3)\times U(3)$ - symmetric four-quark
interaction and 'tHooft determinant \cite{'tHooft:1976up} which contains the six-quark
interaction
\begin{eqnarray}
\mathcal{L}& =& {\bar q}(i{\hat \partial} - m^0)q
 + \frac{G_1}{2}\sum_{i=0}^8 \left[({\bar q}{\lambda}_i q)^2 +({\bar q}i{\gamma}_5{\lambda}_i q)^2\right]
 - \frac{G_2}{2}\sum_{i=0}^8 ({\bar q}\gamma_\mu{\lambda}_i q)^2 -\nonumber\\
&&- K \left\{ {\det}[{\bar q}(1+\gamma_5)q]+{\det}[{\bar q}(1-\gamma_5)q] \right\}
\label{Ldet},
\end{eqnarray}
where $\lambda_i$ (i=1,...,8) are the Gell-Mann matrices and $\lambda^0 =
{\sqrt{2\over 3}}${\bf 1}, with {\bf 1} being the unit matrix;
$\bar{q}=\{\bar{u},\bar{d},\bar{s}\}$ are antiquark fields and $q$ are quark fields;
$m^0$ is a current quark mass matrix with diagonal elements $m^0_u$, $m^0_d$, $m^0_s$
$(m^0_u \approx m^0_d)$; $G_1$ and $G_2$ are the four-quark coupling constants in the
scalar-pseudoscalar and vector channels; $K$ is the six-quark coupling constant.

In this model a very important role is played by gap equations which describe spontaneous
breaking of chiral symmetry, and connect current and constituent quark masses. These equations
have the form
\begin{eqnarray}
m_u&=&m_u^0 + 8 m_u G I_1^\Lambda(m_u)+32 m_u m_s K I_1^\Lambda(m_u) I_1^\Lambda(m_s),\nonumber\\
m_s&=&m_s^0 + 8 m_s G I_1^\Lambda(m_s)+32  K \left(m_uI_1^\Lambda(m_u)\right)^2, \label{gap}
\end{eqnarray}
where $m_u$ and $m_s$ are the constituent quark masses; $I_1^\Lambda(m)$ is the quadratically
divergent integral describing the quark loop with one vertex, tadpole, with quark mass $m$ and
the cut-off parameter $\Lambda$.

It is possible to use different methods for investigation of the meson behavior in the
hot and dense matter. The most popular one is the Matsubara technique
\cite{Kapusta:1989tk}. However, for many applications it is more convenient to use an
equivalent representation for the quark propagator derived in the "real time"
formalism \cite{Dolan:1973qd,Ebert:1992ag}\\
\begin{eqnarray}
S(p,T,\mu)&=&(\hat{p}+m)\left[ \frac{1}{p^2-m^2+i\epsilon}+\right.\nonumber\\
&&\left.i 2\pi
\delta(p^2-m^2)(\theta(p^0)n(\mathbf{p},\mu)+\theta(-p^0)n(\mathbf{p},-\mu)) \right],
\end{eqnarray}
where
\begin{eqnarray}
n(\mathbf{p},\mu)=\left(1+\exp\frac{E-\mu}{T}\right)^{-1}
\end{eqnarray}
is the Fermi-Dirac function for quarks, $E=\sqrt{\mathbf{p}^2+m^2}$, $T$ is the
temperature and $\mu$ is the chemical potential. Here it is convenient to use the
three-momentum cut-off $\Lambda_3$. As a result, quadratically and logarithmically
divergent integrals $I_1^{\Lambda_3}(m,T,\mu)$ and $I_2^{\Lambda_3}(m,T,\mu)$ take the
form
\begin{eqnarray}
I_1^{\Lambda_3}(m,T,\mu)&=& \frac{N_c}{(2\pi)^2}\int \limits_0^{\Lambda_3} dp
\frac{p^2}{E}\left(1-n(\mathbf{p},\mu)-n(\mathbf{p},-\mu)\right),\nonumber\\
I_2^{\Lambda_3}(m,T,\mu)&= &\frac{N_c}{2(2\pi)^2}\int \limits_0^{\Lambda_3} dp
\frac{p^2}{E^3}\left(1-n(\mathbf{p},\mu)-n(\mathbf{p},-\mu)\right).
\end{eqnarray}
We assume that the model parameters $G_1$, $G_2$, $K$, $m_u^0$, $m_s^0$ and $\Lambda_3$ do not
depend on $T$ and $\mu$ \cite{Ebert:1992ag}. The dependence of constituent quark masses $m_u$,
$m_s$ on $T$ and $\mu$ are calculated from the gap equations (\ref{gap}).

The model parameters in vacuum are fixed with the help of the following physical
quantities: masses of $\pi$, $K$ and $\rho$ mesons, weak pion decay constant $f_\pi$,
strong decay width of $\rho$-meson and mass difference of $\eta$-$\eta^\prime$ mesons.
As a result, model parameters have the following values: $m_u$ = 280 MeV, $m_u^0$ =
2.1 MeV,  $m_s$ = 416 MeV, $m_s^0$ = $51$ MeV, $G_1$ = $3.2$ GeV$^{-2}$, $G_2$ = $16$
GeV$^{-2}$, $K = 4.6$ GeV$^{-5}$, $\Lambda_3 = 1.03$ GeV.

\begin{figure}[tbp]
\begin{center}
\resizebox{0.4\textwidth}{!}{\includegraphics{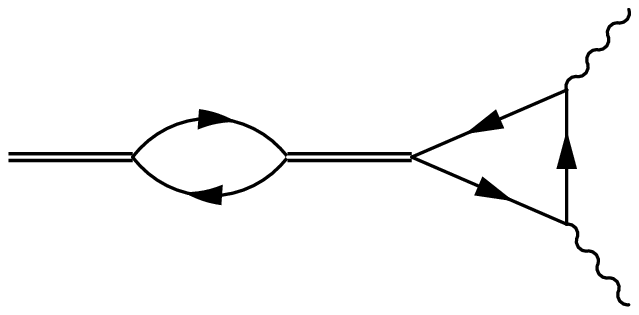}}
\end{center}
\caption{Diagram describing the process $\rho \to \gamma \gamma$. Crossed diagram is not
shown.}\label{Fig_rhogg}
\end{figure}

The amplitude of the process $\rho \to \gamma \gamma $ has the form
\begin{eqnarray}
A^{\alpha \mu \nu}_{\rho \to \gamma \gamma} = J^\alpha_{\rho \to a_0} D_{a_0} T^{\mu\nu}_{a_0
\to \gamma \gamma},
\end{eqnarray}
where $J^\alpha_{\rho \to a_0}$ describe the $\rho \to a_0$ transition, $D_{a_0}$ is
the $a_0$-meson propagator, and $T^{\mu\nu}_{a_0 \to \gamma \gamma}$ is the amplitude
of the $a_0$ two-photon decay (see fig.\ref{Fig_rhogg}).

\begin{figure}[tbp]
\begin{center}
\resizebox{0.49\textwidth}{!}{\includegraphics{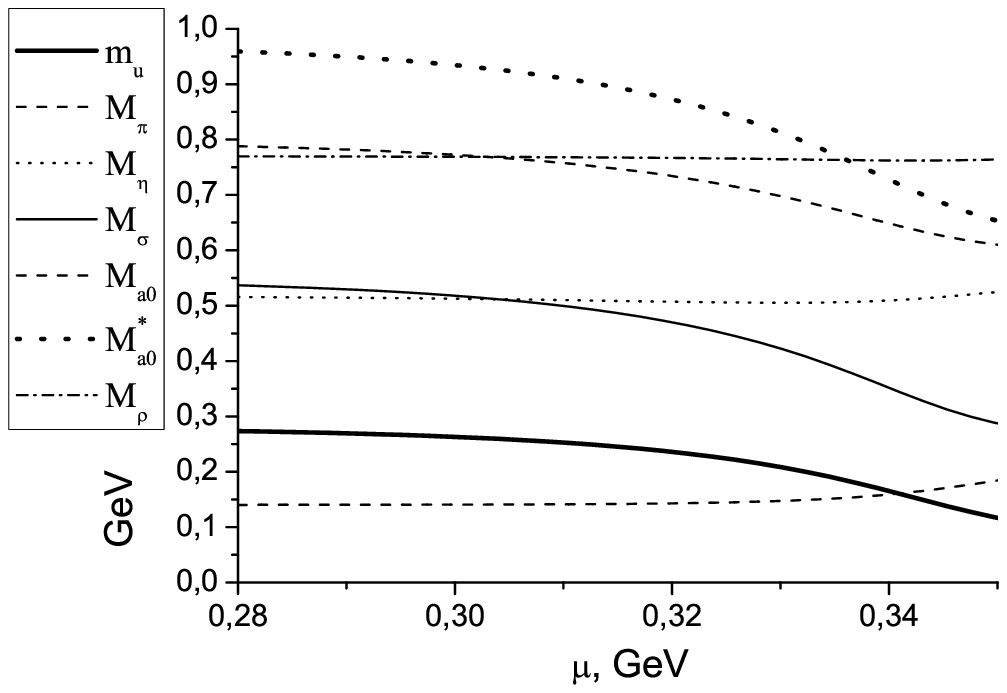}}
\resizebox{0.49\textwidth}{!}{\includegraphics{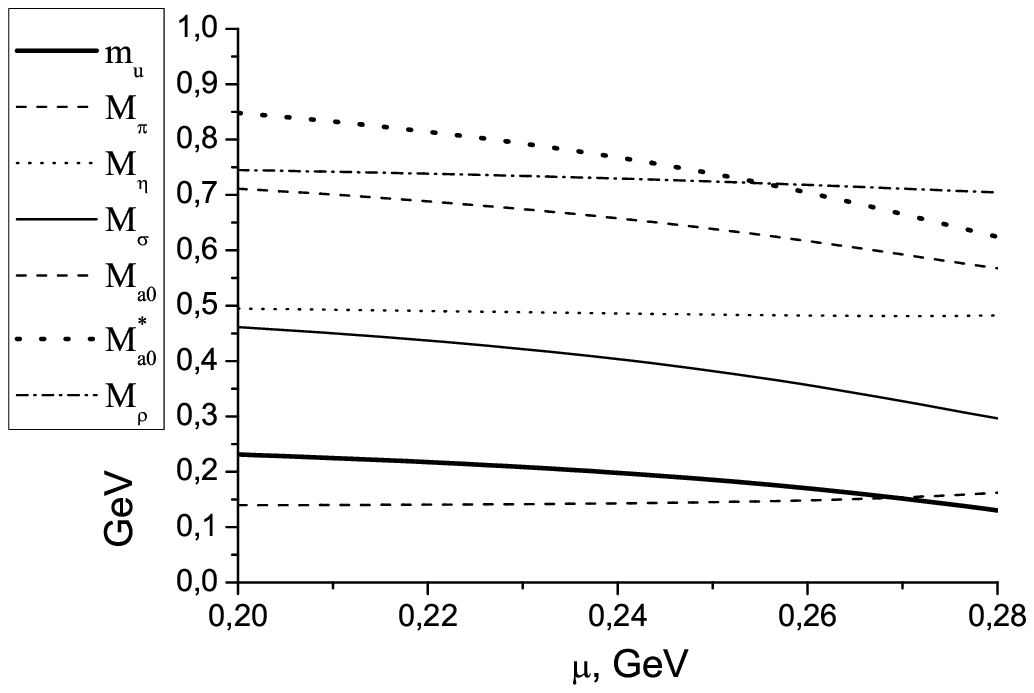}}
\end{center}
\caption{The behavior of the quark mass and meson masses $M_{\pi}$, $M_{a_0}$,
$M_{a_0}^*$, $M_{\rho}$, $M_{\eta}$, $M_{\sigma}$ as function of $\mu$ at
$T=20$(left), $100$(right) MeV.}\label{Fig_masses}
\end{figure}

Let us consider these elements in detail. The transition loop $\rho \to a_0$ has the
form (see \cite{Blaschke:1997jj})
\begin{eqnarray}
J^\alpha_{\rho \to a_0}&=&C_{\rho \to a_0} \frac{1}{4}\int \frac{d^4 k}{(2\pi)^4}
\mathrm{tr_s}\left[S(k_-,T,\mu)\gamma^\alpha S(k_+,T,\mu)\right]=\label{Cra0}\\
&&C_{\rho \to a_0} 2m_u\int \frac{d^4 k}{(2\pi)^3}k^\alpha
\frac{\delta(k_+^2-m^2)}{k_-^2-m^2}
 \left(n(\mathbf{k}_+,\mu)-n(\mathbf{k}_+,-\mu)\right)
 \left(\theta(k_+^0)-\theta(-k_+^0)\right),\nonumber\\
&&C_{\rho \to a_0}=4 N_c g_{a0} g_\rho,\,  k_\pm=k\pm\frac{p}{2},\,
g_{a_0}=(4I_2^{\Lambda_3}(m,\mu,T))^{-1/2},\,
 g_\rho=\sqrt{6}g_{a_0},\nonumber
\end{eqnarray}
where $g_{a_0}$ and $g_\rho$ are the meson-quark coupling constants.

It is easy to see from eq. (\ref{Cra0}) that in vacuum, $\mu=0$, this transition is
forbidden. In medium this transition is nonzero and transversal over $p$, $p_\alpha
J^\alpha_{\rho \to a_0}=0$. Convolution of $J^\alpha_{\rho \to a_0}$ can be expressed
in the form
\begin{eqnarray}
|J|^2=J^\alpha_{\rho \to a_0}J_{\alpha,\rho \to
a_0}=-\frac{p^2}{{\mathbf{p}}^2}J^0_{\rho \to a_0}J_{0,\rho \to a_0} \label{contr}
\end{eqnarray}
Only zero-component of $J^\alpha_{\rho \to a_0}$ is needed (see also \cite{Blaschke:1997jj})
\begin{eqnarray}
J^0_{\rho \to a_0}=\frac{C_{\rho \to a_0} m}{(4\pi)^2|\mathbf{p}|}\int\limits_m^\infty
d E \delta(E)
 \left[ \left(2E+p^0\right) \ln\left(F_+\right) + \left(2E-p^0\right)  \ln\left(F_-\right)
          \right]\nonumber \\
 \delta(E) =
 \frac{\mathrm{sinh}({\mu}/{T})}{\mathrm{cosh}({\mu}/{T})+\mathrm{cosh}({E}/{T})},\,
 F_\pm=\frac{p^2\pm2p_0E+2|\mathbf{p}|k}{p^2\pm2p_0E-2|\mathbf{p}|k},
\end{eqnarray}
where $k=\sqrt{E^2-m^2}$.

The propagator of $a_0$-meson has the form
\begin{eqnarray}
D_{a_0} = \frac{1}{M_{a_0}^2-M_\rho^2-i \Gamma_{a_0}(M_\rho)M_{a_0}}\nonumber
\end{eqnarray}
Note, $\rho$-meson mass weakly changes with $T$ and $\mu$ \cite{Ebert:1992ag}(see also
fig. \ref{Fig_masses}). In the NJL model the mass of $a_0$-meson is expressed as
\cite{Volkov:1998ax}
\begin{eqnarray}
{M^2_{a_0}}=g^2_{a_0}\left[\frac{1}{G_{a_0}} - 8I^{\Lambda_3}_1(m_u)\right] + 4m^2_u \nonumber\\
G_{a_0}=G_1 - 4Km_sI_1^{\Lambda_3}(m_s)
\end{eqnarray}
As a result, the $a_0$-meson mass is lower than experimental
\begin{eqnarray}
M_{a_0}\approx 800\,{\mathrm{MeV}},\quad M_{a_0}^{\mathrm{exp}} = 984.7\pm1.2.
\end{eqnarray}
It is possible that $a_0$-meson has an additional four-quark component
\cite{Jaffe:1976ig,Achasov:2003aa,Gerasimov:2004kq}. In order to take it into account
we introduce an additional term $\Delta$ in the mass formula for $a_0$-meson
\begin{eqnarray}
{M_{a_0}^*}^2=M_{a_0}^2 + \Delta \nonumber
\end{eqnarray}
The model part of the mass of $a_0$-meson decreases with increasing $T$ and $\mu$ and
decreasing of order parameter $m_u$. It is natural to suppose that the additional term
$\Delta$ also has a similar behavior. Particularly, we take $\Delta$ proportional to
the square of the order parameter $m_u$, $\Delta \approx 4 m_u^2$.

The decay width $a_0 \to \eta \pi$ is equal to
\begin{eqnarray}
\Gamma_{a_0\eta\pi}(M_\rho)&=& \frac{g_{a_0\eta\pi}^2}{16 \pi M_\rho}
\sqrt{\left(1-\left[\frac{M_\eta+M_\pi}{M_\rho}\right]^2\right)
\left(1-\left[\frac{M_\eta-M_\pi}{M_\rho}\right]^2\right)}\nonumber\\
g_{a_0\eta\pi}&=& 2m_u g_{a_0} \mathrm{sin}\bar{\theta}, \quad {\bar \theta}=\theta -
\theta_0,
\end{eqnarray}
where $\theta_0 \approx 35.3^{\circ}$ is the ideal mixing angle $({\rm ctg}~ \theta_0={\sqrt
2})$ and $\theta$ is the singlet-octet mixing angle for pseudoscalar mesons.

The amplitude of the two-photon decay of $a_0$ meson has the form \cite{Volkov:1994eu}
\begin{eqnarray}
T^{\mu\nu}_{a_0\to\gamma\gamma} =  C_{a_0\to \gamma \gamma}f_1(T,\mu)F^{\mu\nu},\quad
C_{a_0\to \gamma \gamma}=\frac{2 \alpha g_{a_0}}{3\pi m}\nonumber,
\end{eqnarray}
where $F^{\mu\nu}$ is the electromagnetic tensor, and function $f_1$ is
\cite{Volkov:1997dx}
\begin{eqnarray} \label{f1} f_1(T,\mu)&=&1-\frac{3}{2}m^2(T, \mu)\int_0^\infty dk \frac{k^3}{E^6(k)}
\ln\bigg[\frac{E(k)+k}{E(k)-k}\bigg] \big[ n(k;T,\mu)+n(k;T,-\mu)\big] \nonumber
\end{eqnarray}

\begin{figure}[f]
\begin{center}
\resizebox{0.49\textwidth}{!}{\includegraphics{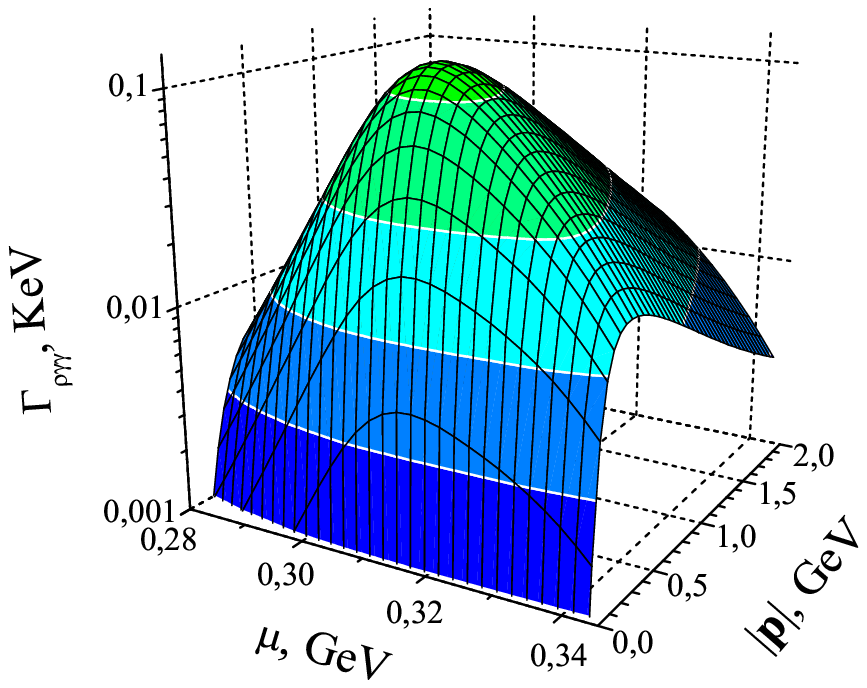}}
\resizebox{0.49\textwidth}{!}{\includegraphics{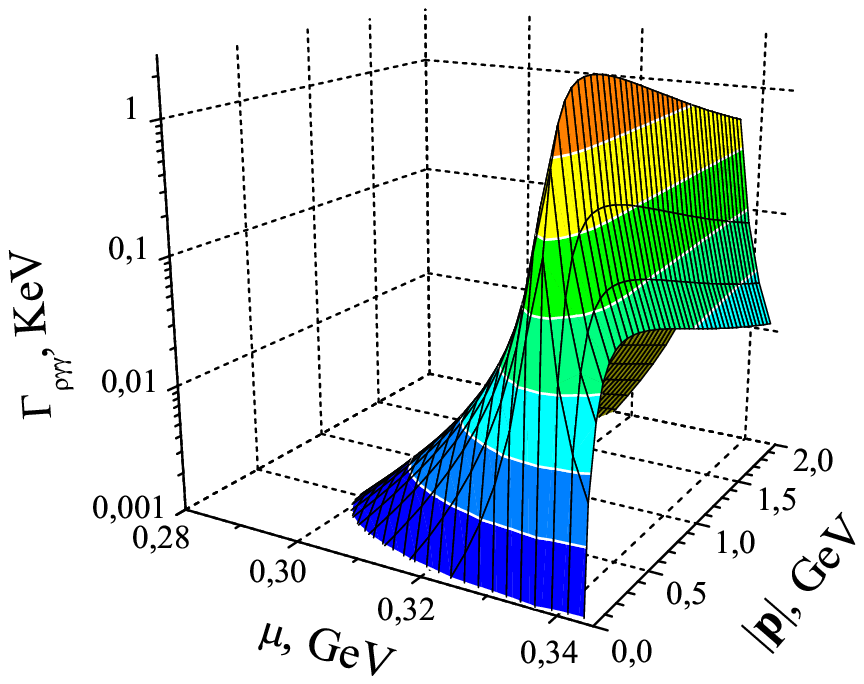}}
\end{center}
\caption{Two-photon decay width of $\rho$-meson with model mass of $a_0$ meson
$M_{a_0}$(left) and with corrected mass of $a_0$ meson ${M_{a_0}^*}$(right) as a
function of $\mu$ and $|\mathbf{p}|$ for $T=20$ MeV.}\label{Fig_Ma0}
\end{figure}

\begin{figure}[f]
\begin{center}
\resizebox{0.49\textwidth}{!}{\includegraphics{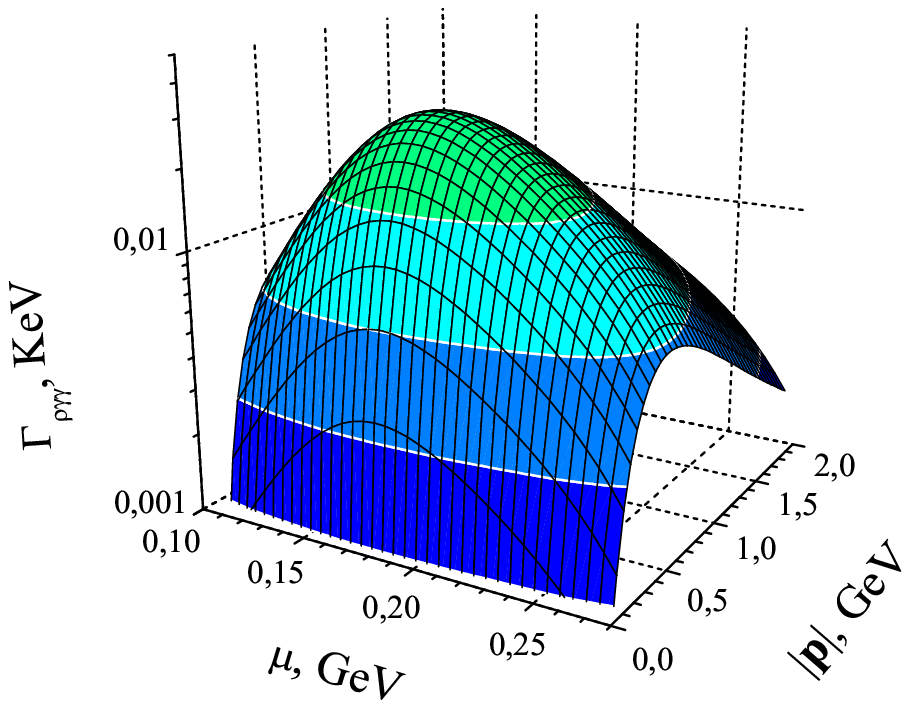}}
\resizebox{0.49\textwidth}{!}{\includegraphics{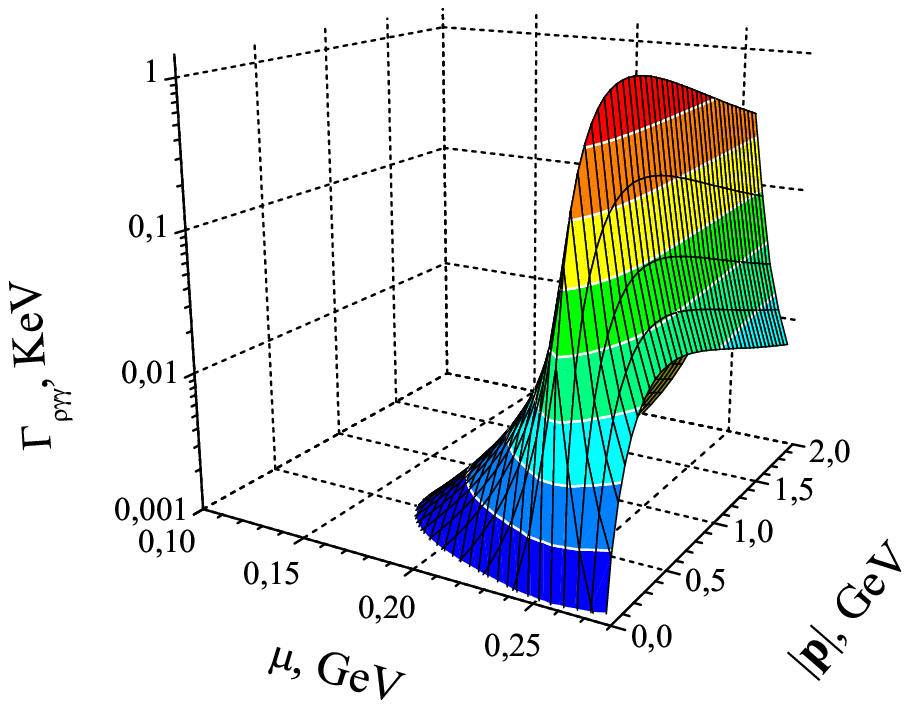}}
\end{center}
\caption{Two-photon decay width of $\rho$-meson with model mass of $a_0$ meson
$M_{a_0}$(left) and with corrected mass of $a_0$ meson ${M_{a_0}^*}$(right) as a
function of $\mu$ and $|\mathbf{p}|$ for $T=100$ MeV.}\label{Fig_Ma0m}
\end{figure}

Now we can consider a full amplitude of the decay $\rho \to \gamma \gamma$. It is
shown in figs.\ref{Fig_Ma0}, \ref{Fig_Ma0m} for temperatures $T=20$, $100$ MeV as
functions of $\mathbf{p}$, $\mu$. It is easy to see that at some values of $T$ and
$\mu$ the two-photon decay width of $\rho$ meson can be comparable with the two-photon
decay width of scalar and pseudoscalar mesons in vacuum (see table \ref{tabu}).
\begin{table}
\begin{center}
\begin{tabular}[]{|c|c|c|c|c|c|c|}
\hline
Particle  & Mass, MeV&$\Gamma_{\gamma \gamma}$, KeV  \\
\hline
$\pi^0$ &$134.9766\pm0.0006$& $(7.8\pm0.5)\cdot 10^{-3}$ \\
\hline
$\eta$&$547.75\pm0.12$&$1.29\pm0.07$ \\
\hline
$\eta^\prime$&$957.78\pm0.14$&$4.29\pm0.15$ \\
\hline
$\sigma$ &$400-1200$& a few \\
\hline
$a_0$&$984.7\pm1.2$&$0.3\pm0.1$ \\
\hline
$f_0$&$980\pm10$&$0.39\frac{+0.1}{-0.13}$ \\
\hline
\end{tabular}
\end{center} \caption{Two-photon decays of scalar and pseudoscalar mesons in vacuum
\cite{Eidelman:2004wy}.}\label{tabu}
\end{table}

\begin{figure}[f]
\begin{center}
\resizebox{0.49\textwidth}{!}{\includegraphics{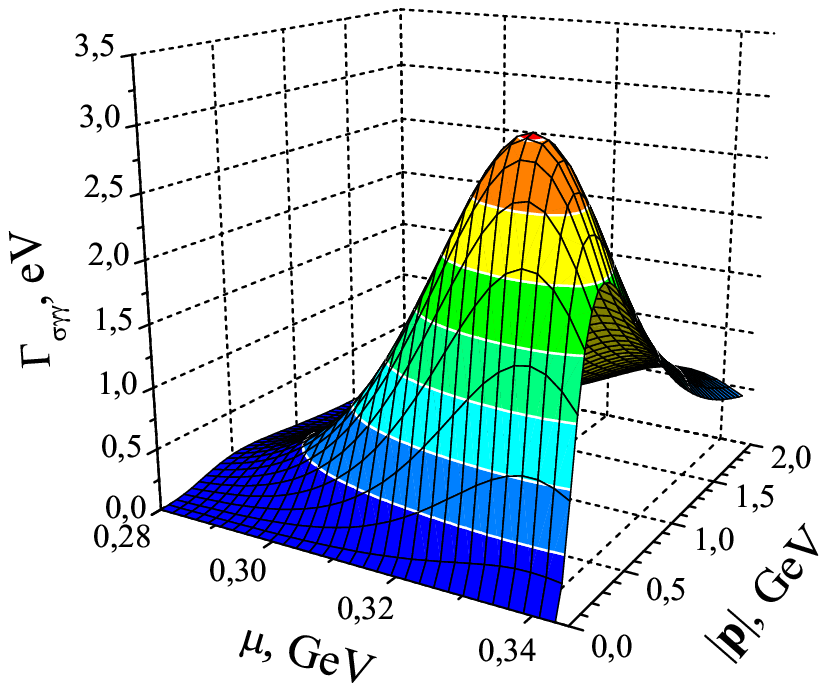}}
\resizebox{0.49\textwidth}{!}{\includegraphics{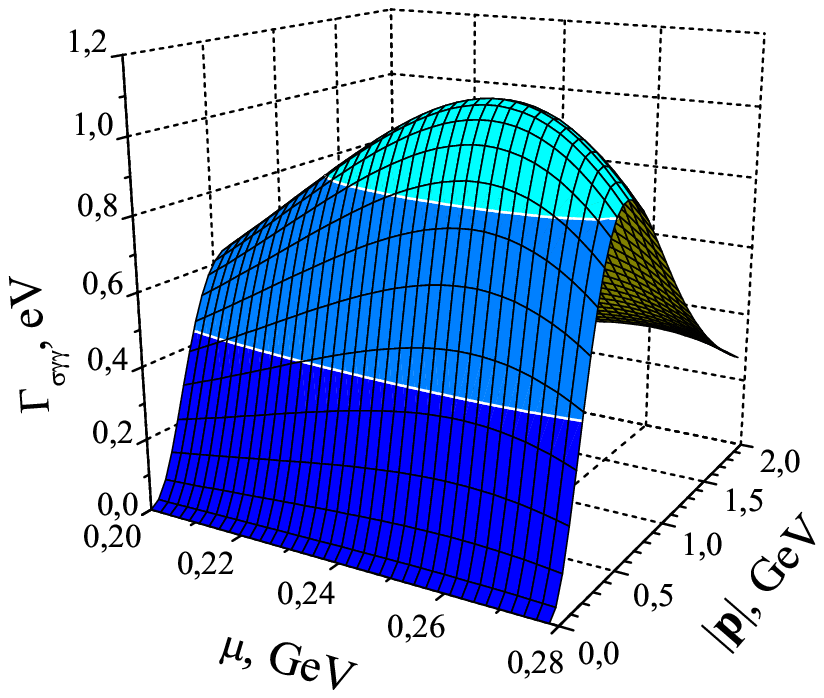}}
\end{center}
\caption{Two-photon decay width of $\omega$-meson as a function of $\mu$ and
$|\mathbf{p}|$ for $T=20$(left), $100$(right) MeV.}\label{Fig_Ms}
\end{figure}

Let us consider the process $\omega \to \gamma \gamma$. The diagrams for this process
consist of $\omega \to \sigma$ transition, $\sigma$-meson propagator and two-photon
decay of $\sigma$ meson. The transition $\omega \to \sigma$ coincides with the
transition $\rho \to a_0$. At large values of $\mu$ and $T$ the $\sigma$-meson mass in
NJL model can approximately be expressed in a very simple form
\begin{eqnarray}
M^2_{\sigma}\approx M^2_{\pi} + 4m^2_u.
\end{eqnarray}
The width of $\sigma$ is defined by the strong decay $\sigma \to \pi \pi$. The decay
width can be expressed as
\begin{eqnarray}
\Gamma_{\sigma \pi \pi}(M_\omega)&=& \frac{3}{2}\frac{g_{\sigma \pi \pi}^2}{16 \pi
M_\omega}\sqrt{1-\frac{4M_\pi^2}{M_\omega^2}},\, \,g_{\sigma \pi \pi}= 2m_u g_{a_0}
\mathrm{cos}\bar{\phi},\,\, {\bar \phi} = \theta_0 - \phi
\end{eqnarray}
where $\phi$ is the singlet-octet mixing angle for scalar mesons.

The amplitude of the two-photon decay of $\sigma$ meson has the same form as $a_0 \to \gamma
\gamma$. The difference is only in numerical factor, $C_{\sigma\to \gamma \gamma}=\frac{10
\alpha g_{a_0}}{9\pi m}$.

The main difference with the decay $\rho \to \gamma \gamma$ is that the $\sigma$ meson
mass is always lower than the $\omega$ meson mass. In the process $\rho \to \gamma
\gamma$ we deal with the sharp resonance. In the case $\omega \to \gamma \gamma$, with
increasing $T$ and $\mu$ the $\sigma$ meson mass will move away from the $\omega$
mass. Therefore, this $\sigma$ meson propagator has no resonant behavior. As a result,
this process is significantly smaller than $\rho \to \gamma \gamma$ as it shown in
fig. \ref{Fig_Ms}.

\begin{figure}[f]
\begin{center}
\resizebox{0.49\textwidth}{!}{\includegraphics{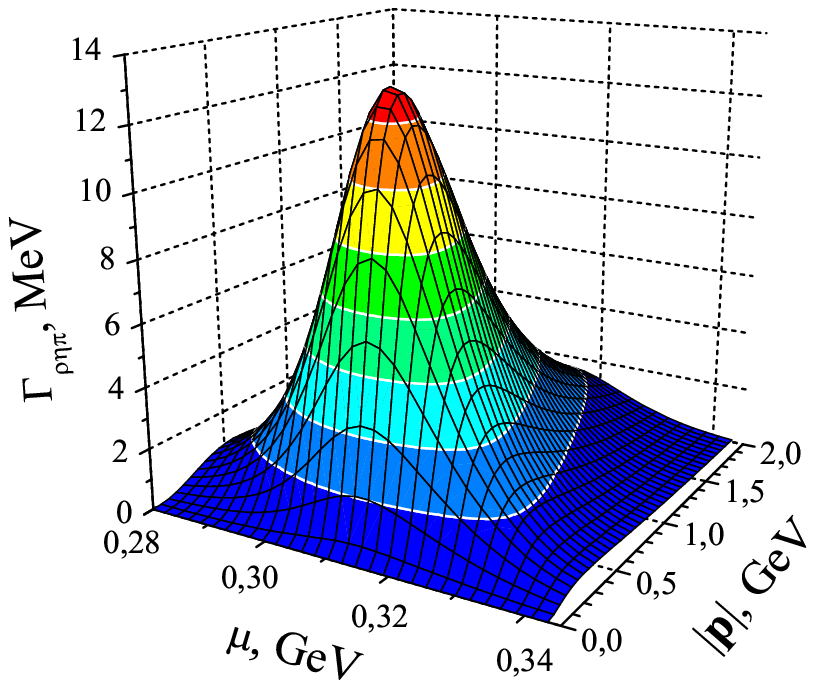}}
\resizebox{0.49\textwidth}{!}{\includegraphics{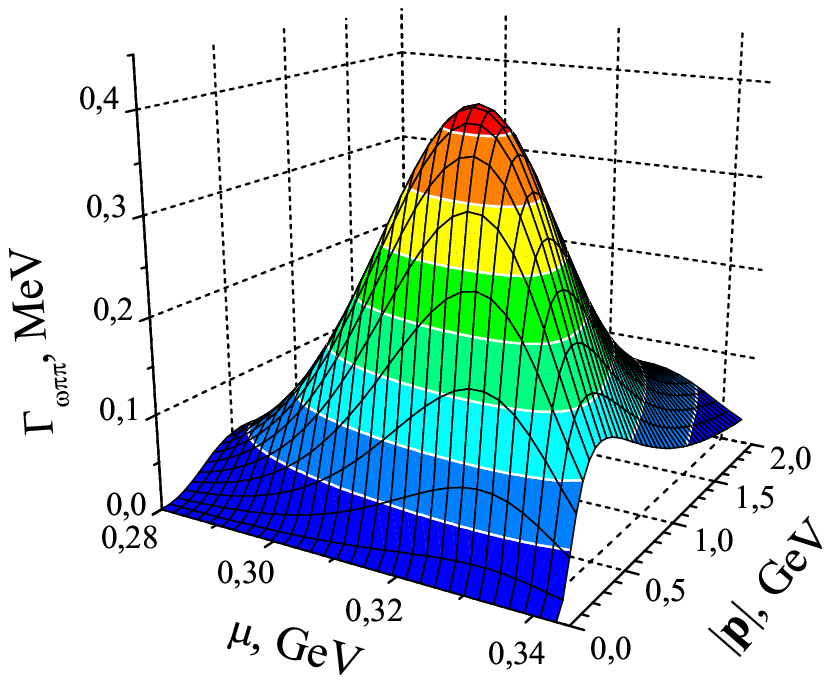}}
\end{center}
\caption{Strong decay widths $\rho \to \pi \eta$ and $\omega \to
\pi \pi$ at temperature $T=20$ MeV.}\label{Fig_strong}
\end{figure}

Strong decays $\rho \to \pi \eta$ and $\omega \to \pi \pi$\footnote{The decay width
$\omega\to\pi\pi$ in medium is calculated in nuclear matter in a hadronic model
including mesons, nucleons and $\Delta$ isobars in \cite{Broniowski:1999ji}.} can be
considered in the same manner. It is necessary to consider the diagrams, which are
similar to those in fig.\ref{Fig_rhogg}, where electromagnetic vertices are changed by
strong ones. These decay widths\footnote{Note that the decay $\rho^0\to \eta \pi^0$ is
closed in vacuum whereas $\Gamma_{\omega\to\pi\pi}=145\pm25$ KeV is only due to
breaking of isotopic invariance \cite{Eidelman:2004wy}.} are given in
fig.\ref{Fig_strong} at temperature $T=20$ MeV.

The main mechanism that opens decays in medium is the quark loop describing a
transition of vector meson into scalar meson. In vacuum this loop is equal to zero.
However, at finite values of the chemical potential and three-momentum this transition
becomes possible \cite{Blaschke:1997jj}.

In spite of this transition, intermediate scalar mesons $\sigma$ and $a_0$ take part
in two-photon vector meson decays. They play different roles in these processes.
Indeed, $a_0$-meson in the region near the critical $\mu$ values becomes a sharp
resonance that leads to noticeable amplifications of decays $\rho\to\gamma\gamma$,
$\rho\to\eta\pi$ whereas $\sigma$-meson taking part in the decays of $\omega$-meson
has other behavior. It has a non-resonance character and does not lead to
amplification of processes. As a result the maximal value of the $\rho\to\gamma\gamma$
decay width at $T=20$ MeV is equal to $2.8$ KeV at $\mu=337$ MeV and
$|\mathbf{p}|=360$ MeV. The maximal value of the $\omega\to\gamma\gamma$ decay width
at $T=20$ MeV is equal to $3.1$ eV at $\mu=330$ MeV and $|\mathbf{p}|=410$ MeV.

Similar resonance effects with $a_0$-meson are considered in the process $\eta \pi \to
e^+ e^-$ \cite{Teodorescu:2000mg}. Note that $\sigma$-meson can also lead to resonance
amplification of different processes where it participates as an intermediate state,
such as $\pi\pi \to \gamma\gamma$ \cite{Volkov:1997dx}, $\pi\pi \to \pi\pi$
\cite{Jido:2000bw} and $\pi\pi \to e^+ e^-$ \cite{Teodorescu:2000mg}.

In this paper, we do not consider nonresonance diagrams, such as a direct decay of
$\rho$ meson into two photon through the quark triangle diagram (this type of diagrams
is considered in \cite{Teryaev:1996dv,Skalozub:1999iw}). The diagrams with pion loops
\cite{Weldon:1991ej} also neglected because they have next order $1/N_c$ expansion. In
future, we plan to consider non-resonance diagrams, process $\phi \to \gamma \gamma$
and dilepton processes with scalar mesons.

The authors should like to thank S.B. Gerasimov, Yu.L. Kalinovsky, O.V. Teryaev, and V.L.
Yudichev for useful discussions. The work is partially supported by RFBR Grant no.
05-02-16699.


\begin{thebibliography}{10}

\bibitem{Lourenco:2001wi}
C. Lourenco,
\newblock Nucl. Phys. A698 (2002) 13, hep-ex/0105053.

\bibitem{Roland:2001me}
C. Roland et~al.,
\newblock Nucl. Phys. A698 (2002) 54, hep-ex/0105043.

\bibitem{Gazdzicki:2004ef}
M. Gazdzicki et~al.,
\newblock J. Phys. G30 (2004) S701, nucl-ex/0403023.

\bibitem{Volkov:1997dx}
M.K. Volkov et~al.,
\newblock Phys. Lett. B424 (1998) 235, hep-ph/9706350.

\bibitem{Jido:2000bw}
D. Jido, T. Hatsuda and T. Kunihiro,
\newblock Phys. Rev. D63 (2001) 011901, hep-ph/0008076.

\bibitem{Volkov:1986zb}
M.K. Volkov,
\newblock Fiz. Elem. Chast. Atom. Yadra 17 (1986) 433.

\bibitem{Klimt:1989pm}
S. Klimt et~al.,
\newblock Nucl. Phys. A516 (1990) 429.

\bibitem{Klevansky:1992qe}
S.P. Klevansky,
\newblock Rev. Mod. Phys. 64 (1992) 649.

\bibitem{Volkov:1993jw}
M.K. Volkov,
\newblock Phys. Part. Nucl. 24 (1993) 35.

\bibitem{Ebert:1994mf}
D. Ebert, H. Reinhardt and M.K. Volkov,
\newblock Prog. Part. Nucl. Phys. 33 (1994) 1.

\bibitem{'tHooft:1976up}
G. 't~Hooft,
\newblock Phys. Rev. Lett. 37 (1976) 8.

\bibitem{Kapusta:1989tk}
J.I. Kapusta,
\newblock Finite Temperature Field Theory (Cambridge Univ. Press., 1999).

\bibitem{Dolan:1973qd}
L. Dolan and R. Jackiw,
\newblock Phys. Rev. D9 (1974) 3320.

\bibitem{Ebert:1992ag}
D. Ebert et~al.,
\newblock Int. J. Mod. Phys. A8 (1993) 1295.

\bibitem{Blaschke:1997jj}
D. Blaschke et~al.,
\newblock Phys. Rev. C57 (1998) 438, nucl-th/9709058.

\bibitem{Volkov:1998ax}
M.K. Volkov, M. Nagy and V.L. Yudichev,
\newblock Nuovo Cim. A112 (1999) 225, hep-ph/9804347.

\bibitem{Jaffe:1976ig}
R.L. Jaffe,
\newblock Phys. Rev. D15 (1977) 267.

\bibitem{Achasov:2003aa}
N.N. Achasov,
\newblock Nucl. Phys. A728 (2003) 425.

\bibitem{Gerasimov:2004kq}
S.B. Gerasimov,
\newblock Nucl. Phys. Proc. Suppl. 126 (2004) 210.

\bibitem{Volkov:1994eu}
M.K. Volkov,
\newblock Theor. Math. Phys. 101 (1994) 1473.

\bibitem{Eidelman:2004wy}
S. Eidelman et~al.,
\newblock Phys. Lett. B592 (2004) 1.

\bibitem{Broniowski:1999ji}
W. Broniowski, W. Florkowski and B. Hiller,
\newblock Eur. Phys. J. A7 (2000) 287, nucl-th/9905040.

\bibitem{Teodorescu:2000mg}
O. Teodorescu, A.K. Dutt-Mazumder and C. Gale,
\newblock Phys. Rev. C63 (2001) 034903, nucl-th/0008006.

\bibitem{Teryaev:1996dv}
O. Teryaev,
\newblock Chin. J. Phys. 34 (1996) 1074.

\bibitem{Skalozub:1999iw}
V.V. Skalozub and A.Y. Tishchenko,
\newblock hep-th/9907097.

\bibitem{Weldon:1991ej}
H.A. Weldon,
\newblock Phys. Lett. B274 (1992) 133.

\end{thebibliography}
\end{document}